\documentclass[a4paper,12pt]{iopart}
 
\usepackage{hyperref}
\usepackage{graphicx}

\usepackage{bm}
 
\usepackage{amsfonts}
\usepackage{dsfont}
\usepackage{psfrag} 
\usepackage{xcolor}

\usepackage{etoolbox}

\makeatletter
\def\@mkboth#1#2{}
\newlength\appendixwidth
\preto\appendix{\addtocontents{toc}{\protect\patchl@section}}
\newcommand{\patchl@section}{%
  \settowidth{\appendixwidth}{\textbf{Appendix }}%
  \addtolength{\appendixwidth}{1.5em}%
  \patchcmd{\l@section}{1.5em}{\appendixwidth}{}{\ddt}%
}
\makeatother

\begin{document}
\bibliographystyle{iopart-num}

\begin{flushright}
	SAGEX-22-10
\end{flushright}

\title[Amplitudes in Fishnet Theories]{The SAGEX Review on Scattering Amplitudes \\
Chapter 9: Integrability of Amplitudes in Fishnet Theories}

\author{Dmitry Chicherin}

\address{LAPTh, Universit\'e Savoie Mont Blanc, CNRS, B.P. 110, F-74941 Annecy-le-Vieux, France}

\author{Gregory P. Korchemsky}

\address{Institut de Physique Th\'eorique\footnote{Unit\'e Mixte de Recherche 3681 du CNRS}, Universit\'e Paris Saclay, CNRS, 91191 Gif-sur-Yvette, France}  

\medskip

\address{Institut des Hautes \'Etudes Scientifiques, 91440 Bures-sur-Yvette, France}

\vspace{10pt}
\begin{indented}
\item[]\today
\end{indented}

\begin{abstract}
We discuss the properties of scattering amplitudes in a conformal bi-scalar fishnet theory that previously appeared in the study of integrable deformations of $\mathcal N=4$ SYM. In distinction with the latter theory, the scattering amplitudes in the conformal fishnet theory do not suffer from infrared divergences and possess enhanced symmetries. We demonstrate that for the simplest four-particle scattering the single- and double-trace partial amplitudes can be computed for arbitrary coupling in the leading color approximation. We also review the invariance of single-trace components of multi-particle fishnet amplitudes in the leading color limit under infinite-dimensional Yangian symmetry.
\end{abstract}

%
%
%
%
%

\newpage 

\tableofcontents

 
\newcommand\lr[1]{{\left({#1}\right)}} 
\newcommand \vev [1] {\langle{#1}\rangle}
\newcommand\re[1]{(\ref{#1})}
\newcommand{\notag}{\nonumber}
\newcommand{\pa}{\partial}

\section{Conformal fishnet theory}

Conformal fishnet theory is a close cousin of a four-dimensional maximally supersymmetric Yang-Mills theory, or $\mathcal N=4$ SYM for brevity. The latter theory has a number of remarkable properties (conformal symmetry, AdS/CFT correspondence) and it is believed to be integrable in the planar limit \cite{Beisert:2010jr}. This opens up an exciting possibility of computing on-shell scattering amplitudes in planar $\mathcal N=4$ SYM for an arbitrary value of 't Hooft coupling constant $\lambda=g_{\rm YM}^2 N_c$. 

To lowest order in the coupling, the scattering amplitudes in planar $\mathcal N=4$ SYM enjoy a dual (super)conformal symmetry \cite{Drummond:2008vq}. Being combined with the conventional (super)conformal symmetry, it can be further promoted to an infinite-dimensional Yangian symmetry \cite{Drummond:2009fd} which is a hallmark of integrability. This symmetry is exact 
at the Born level but it 
becomes anomalous at loop level due to the presence of infrared divergences in on-shell amplitudes. Regularizing these divergences, e.g. by employing the dimensional regularization, one necessarily breaks (part of) the symmetries of the theory. To pursue an integrability approach, one has to determine the corresponding anomalies but this proves to be a nontrivial task \cite{Drummond:2007au,Korchemsky:2009hm,Caron-Huot:2011dec}.

It is natural to ask whether there exist theories in which scattering amplitudes do not suffer from infrared divergences and possess unbroken Yangian symmetry, at the level of loop integrands at least. Conformal fishnet theory belongs to such class of theories.
It describes two interacting complex scalar fields in $d=4$ dimensions and its Lagrangian takes the form
\begin{eqnarray} \nonumber 
\mathcal L = {}&&  N_c \tr\left[ \partial^\mu \bar X \partial_\mu X+\partial^\mu \bar Z \partial_\mu Z 
+ (4\pi \xi)^2 \bar X \bar Z X Z\right] 
\\[2mm] \notag
   {}&& +(4\pi \alpha_1)^2 \left[\tr(X^2) \tr(\bar X^2) + \tr(Z^2) \tr(\bar Z^2) \right]
\\[2mm] \label{L}
{}&& -(4\pi\alpha_2)^2 \left[\tr(XZ) \tr(\bar X\bar Z)+\tr(X\bar Z) \tr(\bar XZ)\right],
\end{eqnarray}
where $X$ and $Z$ are complex $N_c\times N_c$ traceless matrix scalar fields. 

The theory \re{L} naturally arises in the study of 
 integrable deformations of $\mathcal N=4$ SYM, the so-called $\gamma-$deformations \cite{Leigh:1995ep,Lunin:2005jy,Frolov:2005dj}. In this case, one modifies the Lagrangian of $\mathcal N=4$ SYM by inserting the additional phase factors $\e^{\pm i\gamma_j/2}$ in front of Yukawa  and quartic scalar interaction terms, thus introducing the dependence on three parameters $\gamma_{j=1,2,3}$. This modification breaks supersymmetry but preserves integrability of the theory, in the planar limit at least.  One recovers the fishnet theory \re{L} by going to a specific double scaling limit when the Yang-Mills coupling vanishes, $g_{\rm YM}^2\to 0$,
and one of the deformation parameters goes to infinity along imaginary axis $\gamma_3\to\infty$ in such a way that the product 
$\xi^2= g_{\rm YM}^2 N_c \e^{-i\gamma_3}$ remains finite \cite{Gurdogan:2015csr,Kazakov:2018ugh}. In this limit, all fields of $\mathcal N=4$ SYM except two complex scalars get decoupled and the Lagrangian reduces to the expression on first line in \re{L}. 
A somewhat unusual feature of the Lagrangian \re{L} is that the single-trace interaction term  $\tr( \bar X \bar Z X Z)$ is not self-adjoint and, as a consequence, the theory is nonunitary. 
One may also consider the limit when all deformation parameters go to infinity simultaneously $\gamma_{1,2,3}\to i\infty$. In this case, only gauge field decouples 
and the resulting theory describes three complex scalars and four Majorana fermions \cite{Gurdogan:2015csr,Kazakov:2018ugh}. 

\begin{figure}[t!]
\psfrag{X}{$X$}\psfrag{Xb}{$\bar X$}\psfrag{Z}{$Z$}\psfrag{Zb}{$\bar Z$} 
\psfrag{A1}{$\mathcal A_{ {XX\bar X\bar X}}=$}\psfrag{A2}{$\mathcal A_{ {XZ\bar X\bar Z}}=$}
  \begin{center}
 \scalebox{0.9}{\includegraphics{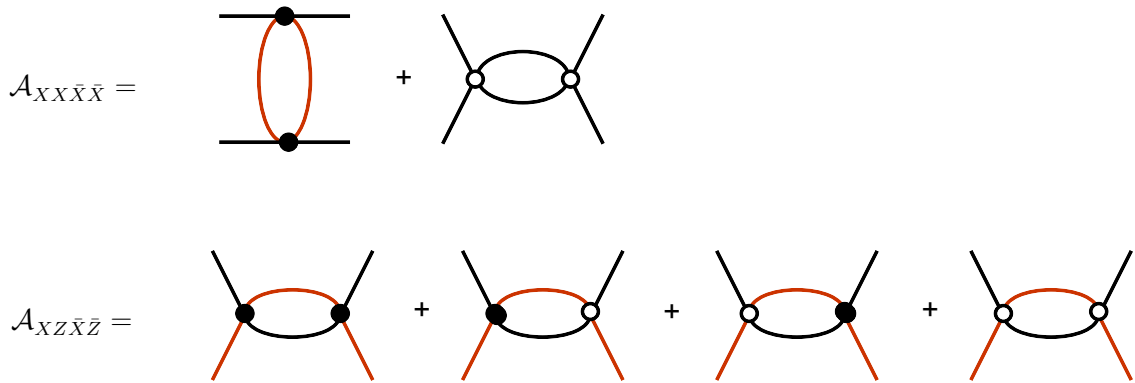}}
    \caption{Four-particle scalar amplitudes at one loop. Black and red lines denote scalars of different flavour. Black and white blobs represent single- and double-trace vertices, respectively. UV divergences cancel in the sum of diagrams at the fixed point.}
    \label{fig:1-loop}
  \end{center}
\end{figure} 

In spite of the fact that $\mathcal N=4$ SYM is an ultraviolet finite theory, the double scaling limit described above generates anomalies and renders the fishnet theory \re{L} ultraviolet (UV) divergent \cite{Fokken:2013aea}. To see this, consider one-loop corrections to the simplest, four scalar scattering amplitudes in the fishnet theory \re{L} coming from Feynman diagrams shown in Figure~\ref{fig:1-loop}. 
Due to a peculiar form of the interaction term in \re{L}, the contribution of the leftmost diagrams containing two single-trace vertices is
proportional a double-trace color tensor  $\tr(T^{a_1} T^{a_2})\tr(T^{a_3}T^{a_4})$ with $T^{a_i}$ being the $SU(N)$ color charges of scalars. In addition, it involves a UV divergent one-loop scalar integral. To make the amplitudes UV finite, one has to 
add to the Lagrangian \re{L} the appropriate counter terms. They are defined on the last two lines of \re{L}.~\footnote{The contribution of other operators like $\tr(X \bar X)\tr(X \bar X)$ is suppressed by powers of $1/N_c$.} 

The counter-terms take the form of double trace operators of canonical dimension $4$.  The corresponding coupling constants $\alpha_1$ and $\alpha_2$ depend on a UV cutoff and develop nontrivial beta functions \cite{Dymarsky:2005uh,Tseytlin:1999ii,Sieg:2016vap,Grabner:2017pgm}. 
In the planar limit, for $N_c\to\infty$ and $\xi^2=$ fixed,
these functions vanish at two complex conjugated fixed points $(\alpha_1^2=\alpha_\star^2,\alpha_2^2=\xi^2)$  and $(\alpha_1^2=\bar\alpha_\star^2,\alpha_2^2=\xi^2)$. In the dimensional regularization, in the $\overline {\rm MS}$ scheme,   one finds \cite{Grabner:2017pgm}
\begin{equation}\label{fixed}
\alpha_\star^2= {i\xi^2\over 2} -{\xi^4\over 2} - {3i\xi^6\over 4} + \xi^8 +  {65i \xi^{10}\over 48} -{19\xi^{12}\over 10} + 
O(\xi^{14})\,. 
\end{equation}
In what follows, we consider only one of these points.

The conformal symmetry of the fishnet theory is restored at the fixed points. Yet another remarkable property of the planar fishnet theory is 
that  it is integrable at the fixed points~\cite{Zamolodchikov:1980mb}. In particular, various four-point correlation functions can be found exactly for an arbitrary value
of the coupling $\xi^2$~\cite{Gromov:2018hut}. We will use this property in the next section to compute four-point scattering amplitudes. 
  
The on-shell amplitudes in the conformal fishnet theory describe the scattering of complex scalars, $X, \bar X, Z, \bar Z$, all belonging to the adjoint representation of the $SU(N_c)$ and carrying a nonzero $U(1)\times U(1)$ charge. 
The scattering amplitudes can be conveniently expanded over the basis of color tensors consisting of single,  double, $\dots$  traces built from the $SU(N_c)$ generators in the fundamental representation, e.g. $\tr(T^{a_1}\dots T^{a_L})$,
$\tr(T^{a_1}\dots T^{a_k})\tr(T^{a_{k+1}}\dots T^{a_L})$ , $\dots$ The leading color contribution only comes from single traces, the contribution of double traces is suppressed by the factor of $1/N_c$.

\begin{figure}[t!]
\psfrag{X}{$X$}\psfrag{Xb}{$\bar X$}\psfrag{Z}{$Z$}\psfrag{Zb}{$\bar Z$}
  \begin{center}
 \scalebox{0.95}{\includegraphics{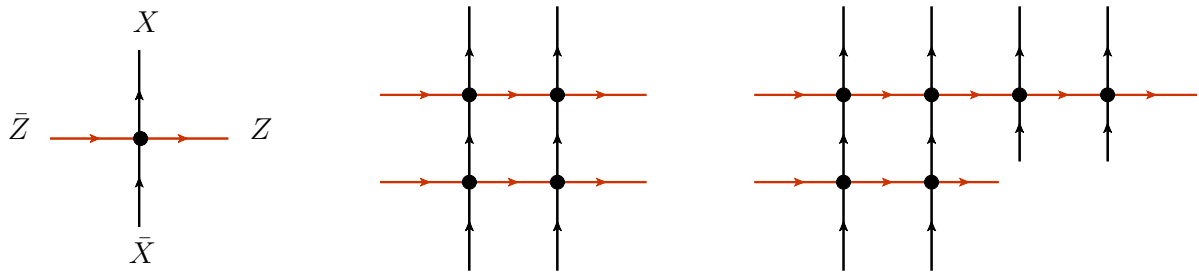}}
    \caption{Planar on-shell scattering amplitudes in fishnet theory. Horizontal (red) and vertical (black) lines denote complex scalars of different flavour.  Black blob represents a single-trace interaction vertex. 
Arrows indicate the charge flow.
    }
    \label{fig:cross}
  \end{center}
\end{figure}

In the planar limit, for $N_c\to\infty$ with $\xi^2$ held fixed,
the scattering amplitudes are given by fishnet diagrams shown in Figure~\ref{fig:cross}. 
These diagrams only involve single-trace interaction vertices and are accompanied by the single-trace color tensors. The diagrams with double-trace interaction vertices are subleading in color and give rise to multi-trace color tensors. 

The simplest, four-particle scattering amplitude is given by the sum of single- and double-trace color tensors.
In ${\cal N} =4$ SYM, the single-trace component is fixed in the leading large $N_c$ limit by a (broken) dual conformal symmetry \cite{Drummond:2007au} and it is given by the ABDK/BDS ansatz \cite{Anastasiou:2003kj,Bern:2005iz} at any loop order. At the same time, the double-trace component has a more complicated form and its all loop representation in ${\cal N} =4$ SYM remains unknown. 
We show below that the four-particle amplitudes are much more under control in the fishnet theory. In spite of the fact that 
 they possess a nontrivial Feynman diagram  expansion, it is  possible to evaluate 
 the single- and double-trace components to all loops to the leading order in $1/N_c$ expansion.

In the next section, we consider four-particle amplitudes in the conformal fishnet theory and present the calculation of the singe- and double-trace partial amplitudes in the leading large $N_c$ limit. In section~\ref{sect3}, we discuss  an infinite-dimensional Yangian symmetry of multi-particle planar amplitudes.

\section{Four particle amplitudes} 
 
The four-particle on-shell scattering amplitude in the conformal fishnet theory \re{L} takes the form 
\begin{eqnarray}\label{co}\notag
\mathcal A = (2\pi)^4 \delta^{(4)}\big(\sum_i  p_i\big) && \left[N_c \tr(T^{a_1} T^{a_2} T^{a_3} T^{a_4})A 
\right.
\\
&&
\left.
+ \tr(T^{a_1} T^{a_2})  \tr(T^{a_3} T^{a_4})  B\right]
+  {\rm perm}\,,
\end{eqnarray}  
where each scalar carries the on-shell momentum $p_i$ (with $p_i^2=0$ and the time component $p_i^0$ being positive or negative for incoming or outgoing particles, respectively) and the color charge $T^{a_i}$ normalized as $\tr(T^a T^b)=\delta^{ab}$. In the case of identical particles, the relation \re{co} contains the additional terms  denoted as `perm' that are needed to restore the Bose symmetry of the amplitude. 

The amplitude \re{co} is different from zero only if the total charge of scattered particles vanishes. As a consequence, the only nontrivial amplitudes are $\mathcal  A(XX\bar X\bar X)$ and $\mathcal A(X Z\bar X \bar Z)$. 
The remaining amplitude $\mathcal A(ZZ\bar Z\bar Z)$ can be obtained from $\mathcal A(XX\bar X\bar X)$ by taking into account the invariance of the Lagrangian \re{L} 
under the transformation $X\to Z^{\rm t}$ and $Z\to X^{\rm t}$, where $X^{\rm t}$ stands for transposed matrix and similar for $Z^{\rm t}$.

\subsection{Single-trace partial amplitude}

The color tensors in \re{co} are accompanied by partial amplitudes $A$ and $B$ which are scalar dimensionless functions 
depending on the Mandelstam invariants $s_{ij}=(p_i+p_j)^2$ and  the coupling constant $\xi^2$.  In addition, they also depend on $1/N_c$. In what follows we consider both functions in the leading large $N_c$ limit, for $N_c\to \infty$ and $\xi^2$ hold fixed. 
 In this limit, the single-trace contribution to \re{co} only comes from planar diagrams containing vertices generated by the  single-trace interaction term in \re{L}. For an arbitrary number of scattering particles they have a form of fishnet diagrams as shown in Figure~\ref{fig:cross}. 
 
In particular, for four particles, the single-trace partial amplitudes are given by their tree-level expressions 
\begin{equation}\label{A-tree}
A_{XZ\bar X\bar Z} = (4\pi \xi)^2 \,, \qquad\qquad A_{XX\bar X \bar X} = 0\,.
\end{equation}
where the subscript indicates the type of scattered particles.
It is instructive to compare this relation with the analogous expression for four-particle amplitudes in planar $\mathcal N=4$ SYM. In this theory, the four-particle amplitudes admit a concise representation as the product of tree-level amplitudes and light-like rectangular Wilson loop  
\begin{equation}\label{BDS}
A_4 = A_{\rm tree} \, \langle{P\exp\lr{ig_{\rm YM} \int_C dx^\mu A_\mu(x)}}\rangle \,,
\end{equation}
where the gauge field is integrated over a null rectangle $C$ built from light-like momenta of scattered particles. We recall that  
the fishnet theory \re{L} arises as a particular limit of $\gamma-$deformed $\mathcal N=4$ SYM when $g_{\rm YM}\to 0 $ and the gauge theory decouples.
The Wilson loop becomes $1$ in this limit and the relation \re{BDS} reduces to \re{A-tree}.

\subsection{Double-trace partial amplitude}

The situation becomes more interesting for the double-trace term in \re{co}. To lowest order in the coupling, the partial waves $B_{XZ\bar X\bar Z}$ and $B_{XX\bar X\bar X}$ are generated by double-trace interaction terms in \re{L} and are given by
$B_{XZ\bar X\bar Z}=-(4\pi \alpha_2)^2$ and $B_{XX\bar X\bar X}=4(4\pi \alpha_1)^2$. At loop level, these functions
receive contribution from both single- and double-trace interaction terms in \re{L} (see Figure~\ref{fig:1-loop}). The diagrams contributing to $B_{XZ\bar X\bar Z}$
and $B_{XX \bar X\bar X}$ involve two different types of double-trace vertices generated by the interaction terms 
$\tr(XZ) \tr(\bar X\bar Z)$ and $\tr(X^2) \tr(\bar X^2)$, respectively. As a consequence,  $B_{XZ\bar X\bar Z}$ depends on the coupling constants $\xi^2$ and $\alpha_2^2$, whereas $B_{XX\bar X\bar X}$ is a function of $\xi^2$ and $\alpha_1^2$. At the fixed point \re{fixed} both functions only depend on $\xi^2$.
At one-loop order, the sum of four diagrams contributing to $B_{XZ\bar X\bar Z}$ (see Figure~\ref{fig:1-loop}) is proportional to $ (\xi^2-\alpha_2^2)^{2}$. At $\ell$ loops, the relevant diagrams contain a chain of $\ell$ scalar loops attached to the external lines, their contribution is proportional to 
$(\xi^2-\alpha_2^2)^{\ell+1}$. As a consequence,  all loop corrections vanish at the fixed point \re{fixed}, for $\alpha_2^2=\xi^2$,  leading to
\begin{figure}[t!] 
  \begin{center}
 \scalebox{0.95}{\includegraphics{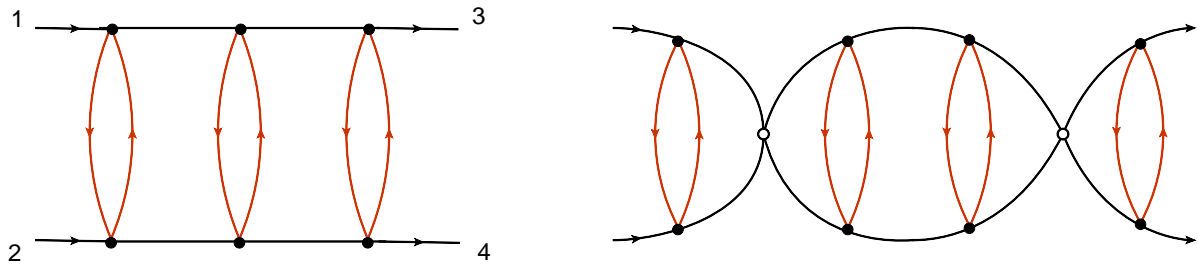}}
    \caption{Two topologies of Feynman diagrams contributing to the scattering amplitude $\mathcal A_{XX\bar X\bar X}$. Notations are the same as in Figures~\ref{fig:1-loop} and \ref{fig:cross}.}
    \label{fig:ladder}
  \end{center}
\end{figure}
\begin{equation}
B_{XZ\bar X\bar Z}=-(4\pi \xi)^2\,.
\end{equation}
Combining this relation with \re{A-tree} we conclude that the single- and double-trace contributions to the scattering amplitude $\mathcal A(X Z\bar X \bar Z)$ are protected from quantum corrections in the planar limit.

The double-trace partial amplitude $B_{XX\bar X\bar X}$ receives a contribution from Feynman diagrams of two different topologies shown in Figure~\ref{fig:ladder}. At the fixed point \re{fixed}, it is a 
dimensionless UV finite function of the ratio of the Mandelstam invariants $s_{13}/s_{12}$  and the coupling constant $\xi^2$. 
One observes that the diagrams  containing at least one double-trace vertex (shown in Figure~\ref{fig:ladder} to the right) naturally factorize into the product of two- and three-point functions. As such they can not depend on the Mandelstam variable
$s_{13}=(p_1+p_3)^2$. The dependence on $s_{13}$ only comes from ladder diagrams shown in Figure~\ref{fig:ladder} to the left. These diagrams have an even number of the single-trace vertices and their contribution runs in powers of $\xi^4$. This suggests to split $B_{XX\bar X\bar X}$ into the sum of two terms that are even and odd in $\xi^2$
\begin{equation}\label{Bpm}
B_{XX\bar X\bar X}=B_-(\xi^2)+B_+(\xi^2,z) +B_+(\xi^2,-z) \,, 
\end{equation}
where $B_-(\xi^2)=-B_-(-\xi^2)$ receives the contribution from the diagrams shown in Figure~\ref{fig:ladder} to the right, whereas $B_+(\xi^2,z)= B_+(-\xi^2,z)$ takes into account the contribution from diagrams of both topologies. 
Here a dimensionless kinematical variable $z$ is related to the scattering angle in the process $1+2\to 3+4$
in the center-of-mass frame 
\begin{equation} 
z=1+ {2s_{13}\over s_{12}}= \cos\theta\,.
\end{equation}
The last term in \re{Bpm} is needed to restore the symmetry of the amplitude under the exchange of particles $1$ and $2$, or equivalently $z\to -z$. 

It is straightforward to compute \re{Bpm} to the first few orders in the coupling \cite{Korchemsky:2018hnb} 
\begin{eqnarray} \label{B-loops}\notag
  B_-(\xi^2) &=& 32i\pi^2\Big[\xi^2 + \xi^6 \left(\frac32+ \frac{\pi
   ^2}3\right)  
   \\
   && 
   + \xi^{10}\left(-\frac{49}{8}+\frac{\pi
   ^2}{6}+\frac{2 \pi ^4}{45}\right) +O\left(\xi^{14}\right)\Big],
 \\[2mm] \notag
  B_+(\xi^2,z)  &=& -16\pi^2 \Big[ \xi^4(H_0 +1) 
 \\ 
 &&
 + \xi^8 \left(H_{-1, 0, 0} + {\pi^2\over 2}  H_{-1} + 4 \zeta_3 - 3\right)   +O\left(\xi^{12}\right)\Big],
\end{eqnarray}
where $\zeta_n=\sum_{k\ge 1} 1/k^n$ are Riemann zeta values and $H_{a_1,a_2,\dots} \equiv H_{a_1,a_2,\dots}((z-1)/2)$ are the so-called harmonic polylogarithms \cite{Remiddi:1999ew,Maitre:2005uu}. 

The explicit expression of $B_+(\xi^2,z)$ becomes rather lengthy at high orders in the coupling. 
This function has interesting properties for $s_{12}\ll s_{13}$, or equivalently $z\to\infty$, corresponding to  
high-energy (or Regge) limit of the scattering amplitude of the process $1+3\to 2+4$. 

It follows from \re{B-loops} that perturbative corrections to $B_+$ in this limit are enhanced by powers of  $L=\ln(z/2)\to \infty$
\begin{equation}\label{B-R}
B_+(\xi^2,z)=-(4\pi)^2 \left[\xi^4 \lr{L+1} +\xi^{8} \lr{\frac{1}{6} L^3 + {\pi ^2\over 2} L }+ O(\xi^{12}) \right].
\end{equation}
Such terms come from the ladder diagrams shown in Figure~\ref{fig:ladder} to the left after integration over the loop momenta in so-called
 the multi-Regge kinematics. In this kinematical regime, the momenta of the exchanged scalars are strongly ordered along the ladder leading to a significant simplification of the corresponding Feynman integrals. This allows one to resum the leading logarithmic (LL) corrections to \re{B-R} of the form $\xi^2(\xi^2 L)^{2n+1}$ (with $n\ge 0$) to all orders to get
\begin{eqnarray} \notag\label{LL}
B^{\rm LL}_+(\xi^2,z) &=& -(4\pi)^2 \xi^2 \sum_{n\ge 0} {(\xi^2 L)^{2n+1}\over (2n+1)n! (n+1)!}  
\\
&=&- 4\pi ^{3/2} \xi^2 (\xi^2L)^{-3/2} \e^{2\xi^2 L}  +\dots \,,
\end{eqnarray}
where dots denote terms suppressed by powers of $\xi^2L=\xi^2\log(z/2)$.

The exponentiation of leading logarithmic corrections in \re{LL} is not surprising and is in agreement with the Regge theory prediction \cite{Gribov:2003nw}. Namely, the asymptotic behaviour of $B_+$ at high energy (for $z\to\infty$) is governed by the leading Regge trajectory and it is expected to have a general form 
\begin{equation} 
B_+(\xi^2,z)\sim (z/2)^{J_R(\xi^2)}\,.
\end{equation}
Comparing this relation with \re{LL}  one finds that the leading Regge trajectory takes the following form at weak coupling
\begin{equation} 
J_R=2\xi^2+O(\xi^4)\,.
\end{equation} 
The exact expression for $J_R$ that is valid for an arbitrary coupling is presented below, see \re{Jr}. The additional factor of $(\xi^2L)^{-3/2}$ in \re{LL} implies that the leading Regge singularity is a cut  in the complex plane of angular momentum  rather than a pole.

\subsection{Lehmann--Symanzik--Zimmermann reduction}
 
The relations \re{B-loops} and \re{B-R} define the weak coupling expansion of the double-trace amplitude \re{Bpm}
at the fixed point \re{fixed}. We demonstrate below that for arbitrary $\xi^2$ the functions $B_-$ and $B_+$ admit a compact representation. Its derivation takes advantage of infrared finiteness of the amplitude \re{Bpm} and it is based on applying the Lehmann--Symanzik--Zimmerman (LSZ) reduction formula to the
four-point correlation function
\begin{equation}\label{G}
G(x_1,x_2|x_3,x_4) = {1\over N_c^2}\vev{\tr(X(x_1) X(x_2))\tr(\bar X(x_3) \bar X(x_4))}\,.
\end{equation}
In the planar limit, this correlation function is given by the sum of the same diagrams as in Figure~\ref{fig:ladder} with the only difference that the external on-shell legs with momenta $p_i$ are replaced by (off-shell) scalar propagators attached to the external points $x_i$.
Following the LSZ procedure, one can obtain the amplitude \re{Bpm} as the residue of Fourier transform of \re{G} at the four simultaneous poles $p_i^2=0$
\begin{eqnarray}\label{LSZ}\notag
\lim_{p_i^2\to 0}  p_1^2 p_2^2 p_3^2 p_4^2  
\int \prod_{i=1}^4 d^4 x_i \,\e^{i p_i x_i}&&  G(x_1,x_2|x_3,x_4) 
\\
&& = {(2\pi)^4 \delta^{(4)}(\sum_i p_i)} B_{XX\bar X\bar X}.
\end{eqnarray}
The four-point correlation function \re{G} can be computed exactly in the planar limit by making use of the conformal symmetry of the fishnet theory at the critical point \re{fixed} and iterative form of contributing diagrams. 

Let us neglect for the moment the diagrams containing double-trace vertices and consider only ladder diagrams shown in Figure~\ref{fig:ladder} to the left. 
Viewed in the configuration space, these diagrams can be built by gluing together elementary building blocks defined as
\begin{eqnarray}\label{H0}
\psfrag{x1}{$  x_1$}\psfrag{x2}{$  x_2$}\psfrag{x3}{$  x_3$}\psfrag{x4}{$  x_4$}
  H(x_1,x_2; x_3 , x_4) = D(x_{13}) D(x_{24}) D^2 (x_{34}) = \parbox[c]{25mm}{ {\includegraphics[width=25mm]{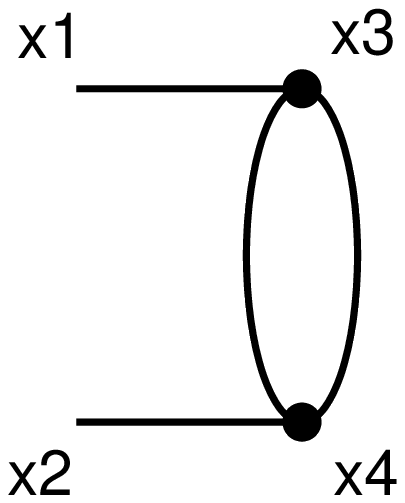}}}
\end{eqnarray}
They are given by the product of four scalar propagators $D(x_{ij}) =-1/(4\pi^2 (x_i-x_i)^2)$.
Namely, introducing an integral `graph generating operator' 
\begin{eqnarray}\label{H}
&& {\bf H}\, f(x_1,x_2) = \int d^4 x_3 \,d^4 x_4\, H(x_1,x_2; x_3, x_4) f(x_3,x_4)\,,
\end{eqnarray}
where $f(x_1,x_2)$ is an arbitrary test function, we can express the contribution of the ladder diagram in Figure~\ref{fig:ladder} as a product of $(n+1)$ copies of the graph generating operators
\begin{eqnarray} \notag \label{geom}
G(x_1,x_2|x_3,x_4) 
 &&=\sum_{n\ge 0} \xi^{4n} (x_1,x_2| \underbrace{\mathbf H \star \dots \star \mathbf H}_{n+1\ \rm times} |x_3,x_4) D^{-2} (x_{34})
\\
&& +(x_1\leftrightarrow x_2) \,,
\end{eqnarray}
where  $ \mathbf H \star \mathbf H$ denotes a convolution of the kernels and the last term is needed to ensure the Bose symmetry of the correlation function.  
The term with $n=0$ yields the Born contribution, $G^{(0)}=D(x_{13}) D(x_{24})+D(x_{14}) D(x_{23}) $. The additional factor of $D^{-2} (x_{34})$ is inserted to compensate the analogous factor in \re{H0}.

Applying \re{geom} we obtain the following all-loop representation of the correlation function  
\begin{eqnarray} \label{G-H}
G(x_1,x_2|x_3,x_4) 
  &&= (x_1,x_2| {\mathbf H \over 1-\xi^4 \mathbf H} |x_3,x_4) D^{-2} (x_{34})+(x_1\leftrightarrow x_2) \,.
\end{eqnarray}
To evaluate this expression one has to diagonalize the integral operator \re{H}. This can be done with a help of conformal symmetry. One can show that the operator $\bf H$ commutes with the generators of conformal  $SO(2,4)$ group in $d=4$ dimensions and, therefore, it is diagonal on the space of functions
 belonging to an irreducible (principal series) representation of this group \cite{Dobrev:1975ru,Dobrev:1977qv}
 \begin{eqnarray}\notag\label{eig}
&& {\bf H} \, f_{\nu,J}(x_1,x_2) = {1\over h(\nu, J)}  f_{\nu,J}(x_1,x_2) \,,
\\[2mm]
&& h(\nu,J) = (\nu^2+J^2/4)(\nu^2+(J+2)^2/4)\,,
\end{eqnarray}
where $J$ is nonnegative integer and $\nu$ real. The function $h(\nu,J)$ takes nonnegative values
and vanishes at $J=\nu=0$.
The eigenfunctions take the form of three-point functions $f_{\nu,J}(x_1,x_2)\sim \vev{\tr(X(x_1) X(x_2)) O_{J,\Delta}(0)}$   where $O_{J,\Delta}$ is a conformal operator of the Lorentz spin $J=0,1,2,\dots$ and the scaling dimension $\Delta=2+2i\nu$. 

Replacing $\bf H$ in \re{G-H} by its expansion over the eigenstates \re{eig}, one obtains the following representation of the correlation function  
\begin{equation}\label{G-OPE}
G(x_1,x_2|x_3,x_4) = \sum_{J\ge 0} \int_{-\infty}^\infty  
{ {d\nu } \,\mu(\nu,J)\over h(\nu,J)-\xi^4}\Pi_{\nu,J}(x_1,x_2|x_3,x_4)
+ (x_1\leftrightarrow x_2)\,,
\end{equation}
where   
$\Pi_{\nu,J}(x_1,x_2|x_3,x_4)$ is the projector onto the eigenstate with
the Lorentz spin $J$ and scaling dimension $\Delta=2+2i\nu$, its explicit expression can be found in \cite{Gromov:2018hut}. 
The kinematical factor  
\begin{equation}
\mu(\nu,J)={ \nu^2 (4\nu^2+(J+1)^2)(J+1) \over 2^{J+4}\pi^7}
\end{equation}
defines the norm of the eigenstates \re{eig}.
The relation \re{G-OPE} resums the contribution of the ladder diagrams shown in Figure~\ref{fig:ladder} to the left. In a close analogy with the scattering amplitudes, the contribution of the remaining diagrams involving at least one double-trace vertex factorizes into the product of two- and three-point functions. As such, it can only modify the integrand in \re{G-OPE} by terms localized at $J=0$ and proportional to $\delta(\nu)$ and its derivatives.
These additional terms are needed to make a weak-coupling expansion of \re{G-OPE} well-defined. More precisely,
the expansion of $J=0$ term in \re{G-OPE} in powers of $\xi^4$ gives rise to divergent integrals involving powers of $1/h(\nu,J=0)\sim 1/\nu^2$. The singularity at $\nu=0$ disappears in the sum of all diagrams shown in Figure~\ref{fig:ladder}. 

At weak coupling, the coefficients of the expansion of \re{G-OPE} in powers of $\xi^4$ can be expressed in terms of single-valued harmonic polylogarithms \cite{Gromov:2018hut}. 
At finite coupling, the integral over $\nu$ in \re{G-OPE} can be evaluated by closing the integration contour to the lower  half-plane, ${\rm Im }\,\nu<0$, and picking up the residues at two poles located at $h(\nu_a,J)=\xi^4$, or equivalently
\begin{eqnarray}\label{scal}\notag
&& \Delta_a=2+2i\nu_a\,,\qquad\qquad 
\\[2mm]
&& (\nu_a^2+J^2/4)(\nu_a^2+(J+2)^2/4)=\xi^4\,,
\end{eqnarray}
where $a=2,4$.
This leads to the following expression for the correlation function \re{G-OPE}, see \cite{Grabner:2017pgm,Gromov:2018hut}
\begin{equation}\label{G-cont}
G(x_1,x_2|x_3,x_4) = {1\over x_{12}^2 x_{34}^2} \sum_{J/2=Z_+} \, \sum_{\Delta=\Delta_2,\Delta_4}C^2_{\Delta,J}(\xi)  \, g_{\Delta,J}(u,v)\,,
\end{equation}
where  $g_{\Delta,J}(u,v)$ are the well-known four-dimensional conformal blocks depending on two cross-ratios, $u=x_{12}^2x_{34}^2/(x_{13}^2x_{24}^2)$ and $v=x_{23}^2x_{14}^2/(x_{13}^2x_{24}^2)$. The explicit expressions for the coefficient functions $C_{\Delta,J}$ can be found in \cite{Grabner:2017pgm,Gromov:2018hut}. The relation \re{G-cont} takes the form of an expansion over conformal partial waves defined by two primary operators that enter the operator product expansion
$\tr[X(x_1)X(x_2)]\sim O_{\Delta_2,J}(x_1)+ O_{\Delta_4,J}(x_1)$. At weak coupling, their  dimensions \re{scal} are $\Delta_2=J+2+O(\xi^4)$ and $\Delta_4=J+4+O(\xi^4)$, so that the operators have twist $2$ and $4$, respectively.
This explains the choice of the subscript in \re{scal}.

 \subsection{Regge trajectories}
 
Having determined the correlation function \re{G-OPE} one can apply the LSZ reduction formula \re{LSZ} to obtain an analogous relation for the scattering amplitude  \cite{Korchemsky:2018hnb}
\begin{equation}\label{B-int}
B_{XX\bar X\bar X}(z,\xi^2)=(2\pi)^8\sum_{J\ge 0} \int_{-\infty}^\infty  {d\nu \, \mu(\nu,J)\over h(\nu,J)-\xi^4}  
\Omega_{\nu,J}(z) 
+ (z\leftrightarrow -z)\,.
\end{equation}
Here the function $\Omega_{\nu,J}(z)$ is obtained from $\Pi_{\nu,J}(x_1,x_2|x_3,x_4)$ by performing a Fourier transform with respect to the external points and taking the residue at the four massless poles. Being a dimensionless function of the Mandelstam invariants $s_{ij}$, it depends on their ratio $z$ defined in \re{Bpm} and has the following parity properties $\Omega_{\nu,J} (z)=(-1)^J \Omega_{\nu,J} (-z)$ and $\Omega_{-\nu,J} (z)=\Omega_{\nu,J} (z)$. In addition, 
$\Omega_{\nu,J}(z)$ is a polynomial of degree $J$ in $z$ of a general form
\begin{eqnarray}\label{Om}\notag
&& \Omega_{\nu,J}(z) =  Q_{\nu,J}z^J + O(z^{J-2})\,,
\\[2mm] 
&& Q_{\nu,J} = (-1)^J \frac{\sinh (2 \pi  \nu )}{2 \pi  \nu } \frac{\Gamma (J-2 i \nu +1) \Gamma (J+2 i \nu +1)}{ \left[\Gamma \left({J}/{2}-i \nu +1\right) \Gamma \left({J}/{2}+i \nu
   +1\right)\right]^2}\,.
\end{eqnarray} 
The subleading expansion coefficients are even functions of $\nu$, their expressions can be found in \cite{Korchemsky:2018hnb}.  
  
As in the previous case, the integral over $\nu$ in \re{B-int} can be evaluated by deforming the integration contour and picking up the residues at the poles located at $h(\nu,J)=\xi^4$
\begin{equation}\label{B-sum}
B_{XX\bar X\bar X}=- 32i \pi^2 \sum_{J/2 =Z_+}\, \sum_{\nu=\nu_2,\nu_4} \frac{2^{1-J} (J+1)\left((J+1)^2+4 \nu ^2\right)\nu }{(J+1)^2+4 \nu ^2+1} \, \Omega_{\nu,J} (z).
\end{equation}
The dependence on the coupling constant enters into this relation through the functions  $\nu_a=\nu_a(J,\xi^2)$ (with $a=2,4$) defined in \re{scal}.  

Let us match the relation \re{B-sum} to the expected expression \re{Bpm} of the scattering amplitude.
It follows from \re{scal} that the weak coupling expansion of $\nu_a(J,\xi^2)$ runs in powers of $\xi^4$ except of $\nu_2(J=0,\xi^2)=-\xi^2+O(\xi^6)$. This means that only one term in the sum \re{B-sum} with $J=0$ and $\nu=\nu_2$ yields the odd function $B_-(\xi^2)$, all remaining terms contribute to the even function $B_+(\xi^2,z)$.
Replacing $ \nu_2(0,\xi^2)=-{\sqrt{\sqrt{4 \xi ^4+1}-1}}/{\sqrt{2}}$ and $\Omega_{\nu,0}(z) = Q_{\nu,0}$ in \re{B-sum} we find 
\begin{equation} \label{B-res}
B_-(\xi^2)= i \frac{32\sqrt{2} \left(1-2 \sqrt{4 \xi ^4+1}\right) }{  \sqrt{4 \xi ^4+1} \sqrt{\sqrt{4 \xi ^4+1}-1}} \sinh ^2\left(\frac{\pi  \sqrt{\sqrt{4 \xi
   ^4+1}-1}}{\sqrt{2}}\right).
\end{equation}  
As expected, $B_-$ is independent of $z$. It is straightforward to verify that its weak coupling expansion agrees with \re{B-loops}. At strong coupling, for $\xi\gg 1$, the amplitude \re{B-res} grows exponentially fast $B_-\sim -64i \e^{\pi\xi}/\xi$.
 
We recall that for $-1< z<1$ the function $B_+(\xi^2,z)$ describes the scattering amplitude of  the process $1+2\to 3+4$.
Replacing $\Omega_{\nu,J} (z)$ in \re{B-sum} by its expansion over Legendre polynomials $\Omega_{\nu,J} (z)=\sum_{k=0}^{J/2} \omega_{k,J,\nu} P_{J-2k}(z)$ one can obtain the standard representation of the scattering amplitude as a sum over partial waves 
\begin{equation}
B_+(\xi^2,z) = \sum_{\ell/2=Z_+} f_\ell(\xi^2)  P_\ell(\cos\theta)\,,
\end{equation}
where $z=\cos\theta$ depends on the scattering angle and $f_\ell(\xi^2)$ depends on the coupling constant through the functions  $\nu_a$ (with $a=2,4$) defined in \re{scal}.  

In the Regge limit, for $z\gg 1$, individual terms in the sum \re{B-sum} grow as a power of $z$ due to $\Omega_{\nu,J}(z) \sim z^J$. 
To find the asymptotic behavior of $B_+(\xi^2,z)$ in this limit, one employs Sommerfeld-Watson transformation to convert the sum over spins in \re{B-int} to a contour integral in a  complex $J-$plane
\begin{eqnarray}\label{A-J}
B_+=  8\pi  \int_C {dJ\over 2\pi i} {\pi \, z^J \over\sin(\pi J)} \int_{-\infty}^\infty d\nu   \frac{2^{1-J} (J+1) \left((J+1)^2+4 \nu ^2\right)\nu ^2\,Q_{\nu,J}  }{
   \left( \nu ^2+ {J^2}/{4}\right) \left(\nu ^2+ (J+2)^2/4\right)-\xi
   ^4 } \,,
\end{eqnarray}
where $Q_{\nu,J}$ is given by \re{Om} and the integration contour $C$ encircles the nonnegative integer $J$ in anti-clockwise direction. The integrand in \re{A-J} has four Regge poles in the complex $J-$plane located at 
\begin{eqnarray}\notag\label{soln}
{}&& J_2^\pm=-1 +\sqrt{1-4 \nu ^2\pm 4 \sqrt{\xi ^4-\nu ^2}} \,,
\\
{}&& J_4^\pm=-1 -\sqrt{1-4 \nu ^2\pm 4 \sqrt{\xi ^4-\nu ^2}} \,.
\end{eqnarray}
They can be interpreted as four branches of the complex curve $h(\nu,J)=\xi^4$ where the function $h(\nu,J)$ is defined in \re{eig}. 

We recall that the scaling dimensions of local operators in the conformal fishnet theory satisfy similar relation \re{scal}. Indeed, the solution to \re{scal} 
define special points on the Regge trajectories located at nonnegative integer $J$ and complex $\nu$. 
The fact that the Regge trajectories describe both the asymptotic behaviour of the scattering amplitudes and the scaling dimensions of local operators is a general feature of conformal theories \cite{Costa:2012cb,Gillioz:2020mdd}. A distinguished feature of the conformal fishnet theory is that the Regge trajectories \re{soln} can be found exactly for arbitrary coupling.

Deforming the integration contour in \re{A-J} we find that each of the Regge poles \re{soln} produces a contribution $B_+ \sim z^{J_a^\pm}$ (with $a=2,4$). The leading behaviour of $B_+$  at large $z$ comes from the poles with the maximal ${\rm Re}  (J_a^\pm)$. 
At weak coupling,  the poles $J=J_4^\pm$ are subleading and the contribution of the poles $J=J_2^\pm$ to \re{A-J} is given by \cite{Korchemsky:2018hnb,DuttaChowdhury:2019jaa}
\begin{eqnarray}\label{A-fin}
{}&& B_+(\xi^2,z)=  \int_{-\xi^2}^{\xi^2} d\nu \, \left[ F(\nu,J_+) (z/2)^{J_+}-F(\nu,J_-)  (z/2)^{J_-}  \right],
\\\notag
{}&& F(\nu,J) =- \frac{32\pi   \nu  \sinh (2 \pi  \nu ) \Gamma (J-2 i \nu +2) \Gamma (J+2 i \nu
   +2)}{\sin(\pi J)\left(J (J+2)+4 \nu ^2\right) [\Gamma \left( {J}/{2}-i \nu +1\right) \Gamma
   \left( {J}/{2}+i \nu +1\right)]^2},
\end{eqnarray}
where $J_\pm\equiv J_2^\pm(\nu)$ is the Regge trajectory \re{soln}. Expanding \re{A-fin} in powers of $\xi^2$ one can reproduce the relation \re{B-R}.  

At finite coupling, the leading contribution to \re{A-J} only comes from the Regge trajectory $J_2^+(\nu)$ for $\nu\to 0$. At small $\nu$ we find from \re{A-J}
\begin{equation}\label{Jr}
B_+(z) \sim \int_0^{\nu_{\rm max}} d\nu \, \nu^2 z^{J_2^+(\nu)}\sim  {z^{J_R}\over (\log z)^{3/2}} \,,
\end{equation}
where $\nu_{\rm max}\ll 1$ and notation was introduced for $J_R=J_2^+(0)=\sqrt{1+4\xi^2}-1$. The relation \re{Jr} defines the asymptotic behaviour of the scattering amplitude in the high-energy limit $z\to\infty$ for arbitrary coupling $\xi^2$.  
At strong coupling,
$J_R=2\xi + O(\xi^0)$ and the amplitude grows as $B_+\sim z^{2\xi}/(\log z)^{3/2}$.

The above analysis can be extended to determine four-point colour ordered scattering amplitudes in the conformal fishnet theory in $d=6$ dimensions~\cite{Bork:2020szm}.  Another approach to computing scattering amplitudes, based on a reformulation of the fishnet theory in twistor space, was developed in \cite{Adamo:2019lor}.
  
\section{Yangian symmetry of the fishnet amplitudes}\label{sect3}
 
In this Section we review the properties of single-trace components of multi-particle fishnet amplitudes in the leading color limit.
These components are of special interest because they are invariant under infinite-dimensional Yangian symmetry, which is one of the manifestations of integrability of the conformal fishnet theory.
 
The leading-color single-trace component of the scattering amplitudes in the conformal fishnet theory is easy to describe. Given an even number $n$ of cyclically ordered scalar particles of different flavor ($X,\bar{X},Z,\bar{Z}$) carrying light-like momenta $ \{ p \} \equiv p_1^\mu,\ldots,p_n^\mu $, there is a unique planar Feynman graph contributing to the amplitude, see e.g. Figure~\ref{fig:cross}. 
This implies that  the perturbative
expansion of the single-trace leading color amplitude contains only one term $O(\xi^{2L})$ where the number of loops $L$ is unambigously fixed.

\begin{figure}[t!]
\begin{center}
\scalebox{0.7}{\includegraphics{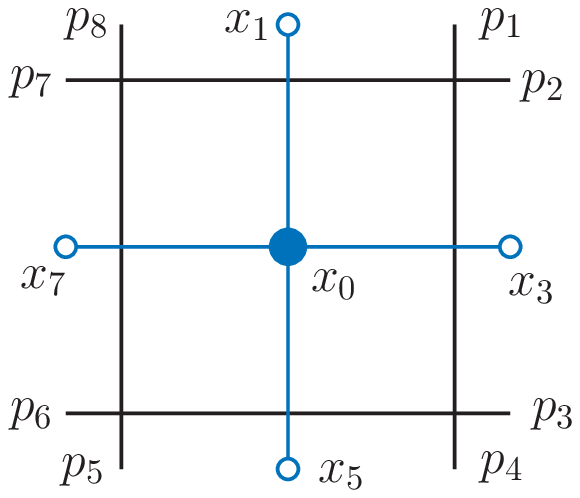}} \qquad
\scalebox{0.7}{\includegraphics{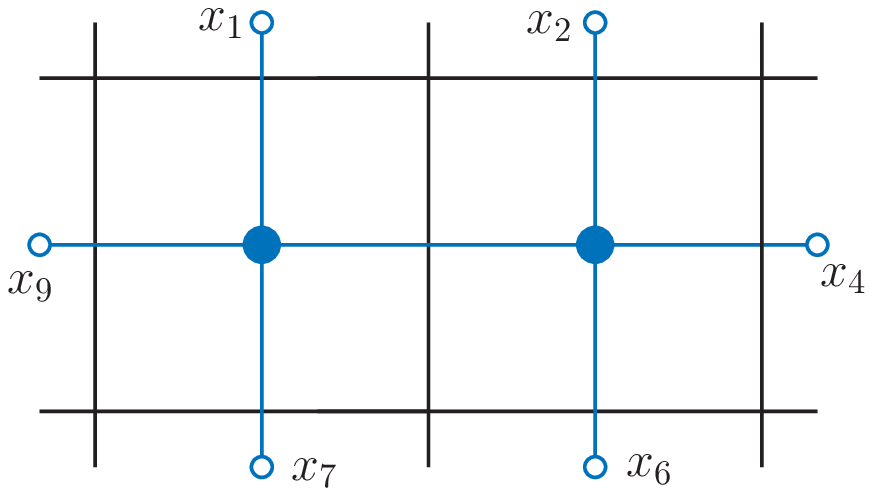}}
\end{center}
\caption{One- and two-loop conformal fishnet graphs in $d=4$ with 8 and 10 external momenta, respectively, are depicted in black. Their dual graphs decorated with dual momenta $x^\mu$ are depicted in blue; $d^4 x$ integrations corresponding to internal vertices are depicted by blue blobs. They are the conformal cross-shape integral $I_{+}$ \re{Icross} and the train track $I_{+\!\!+}$, respectively. }\label{FigCross}
\end{figure}

The fishnet graphs involve only the quartic interaction vertices and, most importantly, they are free of UV and IR divergences. An intuitive way to represent the planar fishnet graphs is to cut them out with scissors from the regular square lattice along a closed line. Some of the multiparticle fishnet amplitudes are given exactly by their tree-level expressions similar to the case of the four-particle amplitude $A_{XZ\bar X \bar Z}$, Eq.~\re{A-tree}. The multiparticle fishnet amplitudes, which are represented by loop graphs, are much more complicated from the analytical point of view.
The prototypical  example is an one-loop box graph which carries 8 external momenta (see Figure~\ref{FigCross}, on the left).
The multiloop graphs contribute only to the scattering amplitudes containing many external particles. Indeed,
cutting out $\ell$ horizontally aligned boxes from the regular lattice, we form  the {\em train track} graphs which are endowed with $2\ell+6$ legs and $\ell$ loops. An example with $\ell=2$ depicted in Figure~\ref{FigCross}, on the right, shows that the two-loop fishnet graph requires at least ten particles. Hence, the multiloop fishnet graphs depend on numerous kinematic invariants. 

This renders their analytic calculation to be an extremely difficult task with little available results. The encountered obstacles are not only technical but also  conceptual since the relevant classes of special functions are not understood well enough.
Indeed,  
the train tracks (for $\ell \geq 2$) evaluate to elliptic functions and are related to the Calabi-Yau geometries \cite{Bourjaily:2018ycu}, not to mention the fishnet amplitudes of generic topology. Therefore, it is even more striking that all fishnet amplitudes are invariant under an infinite-dimensional Yangian symmetry.

This infinite-dimensional symmetry can be formulated in the most intuitive way as the closure of two independent copies of the  conformal group in $d=4$ dimensions. In order to see it, we
represent the particle momenta as differences of the {\em dual momenta}, $p^\mu_i= x^\mu_i -x^\mu_{i+1} $. Then the loop integrations in the fishnet graph $\bar G$ are equivalent to integrations over dual momenta assigned to the internal vertices of the dual fishnet graph $G$. We can also think of the dual fishnet $G$ as representing a correlation function with $n$ external space-time points $ \{ x \}\equiv x_1,\ldots,x_n$. 
Examples of the dual momenta assignment are provided in Figure~\ref{FigCross}; in particular, the one-loop fishnet graph takes the following cross-shape form in the dual momenta,  
\begin{equation}
I_{{\displaystyle +}}( \{ x \} ) = \int \frac{d^4 x_0}{\pi^2} \frac{1}{x_{10}^2 x_{30}^2 x_{50}^2 x_{70}^2} \,, \label{Icross}
\end{equation}
where 
$x_{ij}^2=(x_i-x_j)^2$.

The fishnet amplitude graph $\bar G$ evaluates to a finite function $\bar{I}_{\bar G}$ of $n$ light-like momenta $p^\mu_i$ satisfying on-shell condition $p_i^2=0$ and the momentum conservation,
$
\sum_{i=1}^n p_i^\mu =0 \,.
$
Thus, $\bar{I}_{\bar G}$ is equivalent to a translation-invariant function $I_G$ of $n$ dual momenta $x^\mu_i$ representing the dual fishnet graph $G$, 
\begin{equation}
\bar{I}_{\bar G}(\{ p\}) = I_G(\{ x \}) \,. 
\end{equation}
The quartic interaction vertices and propagators constituting fishnet $\bar{G}$ and its dual $G$ are conformal covariant 
at $d=4$, thus both $I_G$ and $\bar{I}_{\bar{G}}$ are conformal covariant as well. The conformal transformations of the dual fishnet $I_G$ are realized by the first order differential operators in the dual momenta $x^\mu_i$,
\begin{eqnarray}
& D = - i x^\mu \pa_{x^\mu} -i \Delta \,, \quad & L_{\mu \nu} = i x_\mu \pa_{x^\nu} - i x_\nu \pa_{x^\mu} ,  \notag\\[2mm]
& P_\mu = - i \pa_{x^\mu} \,, & K_\mu = i x^2 \pa_{x^\mu} -2i x_\mu x^\nu \pa_{x^\nu} - 2 i x_\mu \Delta \,, \label{confKx}
\end{eqnarray}
with the canonical conformal weights $\Delta$ of the scalar fields in $4d$.
On the other hand, the fishnet amplitude $\bar{I}_{\bar G}$ is annihilated by the momentum space conformal generators which are the second order differential operators in the momenta $p^\mu_i$,
\begin{eqnarray}
& \bar{D} = i p^\mu \pa_{p^\mu} + i (d-\Delta) \,, \quad & \bar{L}_{\mu\nu} = i p_\mu \pa_{p^\nu} - i p_\nu \pa_{p^\mu} \,, \notag \\[2mm]
& \bar{P}_\mu = p_\mu \,, & \bar{K}_\mu = p_\mu (\pa_{p^\nu})^2 - 2 p^\nu \pa_{p^\nu} \pa_{p^\mu} -2(d-\Delta)\pa_{p^\mu} \,. \label{confKp}
\end{eqnarray} 
The two copies of the conformal group are usually referred to in the amplitude context as the dual-conformal and the ordinary conformal symmetries, respectively \cite{Henn:2011xk}. An obvious drawback of this approach is the necessity of dealing simultaneously with two sets of variables. Instead, the Yangian enables us to express both conformal symmetries as differential constraints on $I_G(\{ x \})$ and to elucidate integrability of the fishnet graphs. In the following we rely mostly on the dual momenta variables, and consider the dual fishnet graphs $G$. 

As a matter of fact, the conformal invariance of fishnet amplitudes $\bar{I}_{\bar{G}}$, and hence their Yangian symmetry, takes place only in the off-shell setting, namely for $p_i^2 \neq 0$, which is the situation mostly addressed in the literature. So in the following we assume that all momenta are off-shell. In section \ref{sec_Yang-on-shell} we come back to the on-shell momenta setting.
  
\subsection{Yangian algebra}

The Yangian, as it is defined in the first realization \cite{Drinfeld:1985rx}, is an infinite-dimensional Hopf algebra with a nontrivial co-product. In the case of interest, the level-zero generators $J^A$ of the Yangian are sums of the local conformal generators $J^A_j \in \{ K_{j,\mu}, P_{j,\mu},D_j,L_{j,\mu\nu}\}$ acting on $x_j$, eq. \re{confKx}, and the level-one Yangian generators $\widehat{J}^A $ are bilinear in the local conformal generators,   
\begin{eqnarray} \notag
&& J^A = \sum_{j=1}^n J_j^A \,, \quad
\\
&& \widehat{J}^A(s) = f^{A}{}_{BC} \sum_{1 \leq j < k \leq n} J^C_j J^B_k + \sum_{j=1}^n s_j J_j^A \label{JJhat}\,,
\end{eqnarray}
where $f$'s are structure constants of the $4d$ conformal algebra $so(2,4)$, $A,B,C$ are the corresponding adjoint representation indices, and the numbers $s_j$ are called the evaluation parameters. The remaining higher-level Yangian generators are multiple commutators of $\widehat{J}^A$. A simple graphical rule \cite{Chicherin:2017cns} specifies the {\em evaluation parameters} $s_j = s^{(G)}_j$ for an arbitrary dual fishnet graph $G$ such that the Feynman integral $I_G$ is annihilated by the Yangian generators,  
\begin{equation}
J^A I_G( \{ x \}) = 0  \, ,\qquad
\widehat{J}^A(s^{(G)}) \, I_G( \{ x \}) = 0 \,, \label{JhatInv}
\end{equation}
and invariance with respect to higher Yangian generators follows from \re{JhatInv}.

The Yangian level-one generators $\widehat{J}^A$ in eq.~\re{JJhat} employ a preferred ordering of the dual momenta $x^\mu_j$. We could formulate the Yangian symmetry using as well the cyclically shifted counterparts of \re{JJhat} since the single-trace amplitude is invariant under the cyclic shift. The agreement among cyclic shifts of \re{JhatInv} is insured by the choice of the evaluation parameters.

The Yangian invariance \re{JhatInv} with $s = s^{(G)}$ takes place not only for the Feynman integral $I_G$ with propagators $1/x_{ij}^2$, but also for all its generalized unitarity cuts. The latter can be obtained by replacing one or several propagators with delta-distributions, $1/x_{ij}^{2} \to 2i\pi\delta(x_{ij}^2)$.  

The off-shell box Feynman integral $I_{{\textstyle +}}$, Eq.~\re{Icross}, serves as a simple nontrivial example of the Yangian symmetry application. Taking into account its conformal covariance in the dual momentum space, we can represent it as follows
\begin{equation}
I_{{\textstyle +}}(\{x \}) = \frac{1}{x_{15}^2 x_{37}^2}\phi(u,v) \,, \label{I+}
\end{equation}
where $u = {x_{13}^2 x_{57}^2}/({x_{15}^2 x_{37}^2})$ and $v = {x_{17}^2 x_{35}^2}/({x_{15}^2 x_{37}^2})$ are the cross-ratios. 
The conformal symmetry $J^A$ irreducibly transforms the level-one Yangian generators $\widehat{J}^A$ among each other, so it is sufficient to consider only one of them. The usual choice is $\widehat{P}_\mu$ defined in \re{JJhat} which has the simplest expression among $\widehat{J}^A$'s. 
Thus, imposing $\widehat{P}_\mu-$invariance on the off-shell box in the form \re{I+} we implement all constraints of the Yangian symmetry. This results in the following differential equation \cite{Chicherin:2017frs},
\begin{eqnarray}
u(u-1) \pa^2_u \phi + v^2 \pa_v^2 \phi + 2 u v \pa_u \pa_v \phi + (3u-1)\pa_u \phi + 3 v \pa_v \phi + \phi = 0 \,.
\end{eqnarray}
The latter can be convenitely solved \cite{Loebbert:2019vcj} in the $z,\bar z$ cross-ratio variables. The solutions are the Bloch-Wigner function, as well as its generalized cuts.

The differential constraint implied by the $\widehat{P}_\mu-$invariance is employed as a tool for analytic calculation of the fishnet Feynman integrals dubbed the Yangian bootstrap in \cite{Loebbert:2019vcj}. For a generic fishnet graph it is a system of second order partial differential equations in a number of cross-ratios. 
Despite the fact that solving these differential equations, supplemented with the appropriate boundary conditions, may seem to be an extremely hard task, in the framework of the $d$-dimensional generalization of the bi-scalar fishnet model \cite{Kazakov:2018qbr}, a number of one-loop Feynman integrals have been solved analytically by the Yangian bootstrap \cite{Loebbert:2019vcj,Loebbert:2020glj} as multivariable extensions of the hypergeometric function.

 Both conformal symmetries, \re{confKx} and \re{confKp}, manifest themselves in the first realization of the Yangian as follows,
$\widehat{P}_\mu \sim \bar{K}_\mu$, where $\sim$ implies that the momentum conservation and the conformal weights counting are taken into account. Moreover, we can exchange the two conformal groups, $\widehat{\bar{P}}_\mu \sim K_\mu$, i.e. we can choose the momentum space conformal generators \re{confKp} as the zero-level Yangian generators $\bar{J}^A$, and define the bilocal level-one generators $\widehat{\bar{J}}^A$ according to \re{JJhat}. The Yangian symmetry in this dual formulation of the fishnet graphs $\bar{I}_{\bar G}$  takes place as well,

\begin{equation}
\widehat{\bar{J}}^A(s^{(\bar{G})}) \, \delta^{(4)}\Bigl(\sum_i p_i\Bigr) \bar{I}_{\bar G}(\{ p \} ) = 0 \,. \label{JhatMoment}
\end{equation}

\begin{figure}
\begin{center}
\scalebox{0.8}{\includegraphics{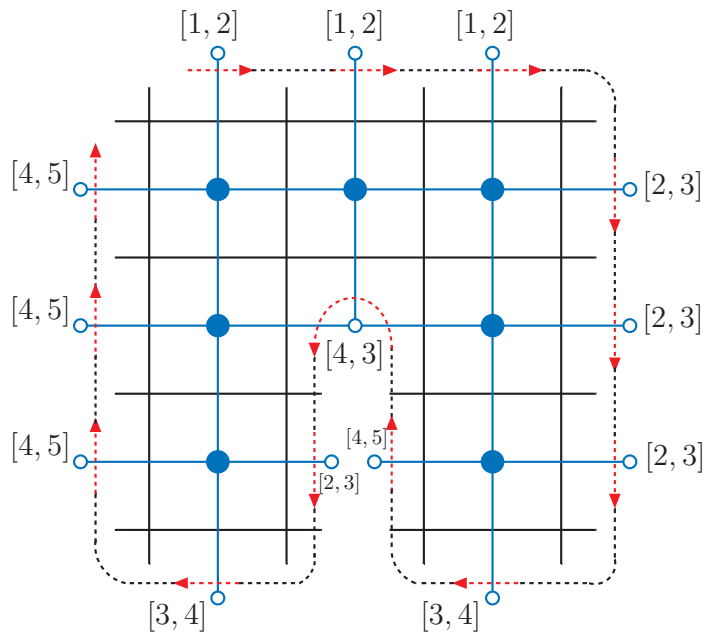}}
\end{center}\caption{A seven-loop $4d$ conformal fishnet graph $\bar{G}$ with 20 external momenta. Its dual fishnet graph $G$ is depicted in blue. The monodromy operator $T(u)$ from \re{Tinv} is represented by the dashed contour encircling the dual graph. The Lax operators $L_i[\delta^+_i,\delta^-_i]$ are depicted with dashed red arrows decorated with parameters $[\delta^+_i,\delta^-_i]$, and the dashed black lines represent their matrix product. The corresponding eigenvalue 
 \re{Tinv} is $\lambda(u) = [3]^3 [4]^7 [5]^4$.}\label{FigPi}
\end{figure}

\subsection{Yangian symmetry from integrable spin chains}

Various formulations of the Yangian symmetry help to highlight its different aspects. The RTT-formulation of the Yangian \cite{Sklyanin:1991ss} is motivated by the quantum integrable spin- chain models. In the context of the fishnet graphs, the quantum states of the relevant spin chain belong to the principal series representation of the conformal group. This non-compact spin chain perspective is very fruitful, and it helps to reveal integrability of the conformal fishnet theory \re{L} at the level of individual Feynman graphs \cite{Gromov:2017cja}. In particular, it allows for evaluating exactly a number of four-point correlation functions \cite{Grabner:2017pgm,Gromov:2018hut} and multi-loop Feynman integrals \cite{Derkachov:2018rot,Derkachov:2021ufp} by solving the spectral problem for the relevant quantum mechanical spin-chain model \cite{Derkachov:2019tzo}. Here we concentrate on the Yangian symmetry implications of the spin-chain perspective.

For the $n$-particle scattering in the conformal fishnet theory, all Yangian generators are packed in the $4\times 4$ matrix $T(u)$ (defined in \re{Tmonod} below)  called the {\em monodromy}  matrix. It is given by a series in the spectral parameter $u$,  
\begin{eqnarray}\notag
&& \frac{1}{u^n}T(u) = \mathds{1} + \frac{1}{u} \mathbb{J} + \frac{1}{u^2} \widehat{\mathbb{J}} + \ldots\,,\;\quad
\\[2mm]
&&
\mathbb{J}_{\alpha\beta}  \equiv J^A  (S_A)_{\alpha\beta} \,,\; \widehat{\mathbb{J}}_{\alpha\beta} \equiv \widehat{J}^A (S_A)_{\alpha\beta} \label{Texp}\,,
\end{eqnarray}
where the level-zero $J^A$ and the level-one $\widehat{J}^A$ generators are contracted with
matrices $(S_A)_{\alpha\beta}$ with $\alpha,\beta=1,\ldots,4$ forming the spinorial representation of $so(2,4)$. All commutation relations among the Yangian generators are encoded in the Yang-Baxter RTT-relation for $T(u)$. 
Besides the spectral parameter $u$, the monodromy also depends on the inhomogeneities $\vec \delta^{\pm} = (\delta^{\pm}_1,\ldots, \delta_n^{\pm}) $ which are analogous to the set of evaluation parameters $s_1,\ldots,s_n$ from \re{JJhat}. Similarly to eq. \re{JhatInv}, a simple graphical rule specifies $\vec\delta^{\pm} = \vec\delta^{\pm}_{(G)}$ for an arbitrary fishnet $I_G$ such that its Yangian invariance takes the form of the eigenvalue equation with eigenvalue $\lambda$,
\begin{equation}
T_{\alpha\beta}\left(u,\vec\delta^{\pm}_{(G)}\right) I_G = \delta_{\alpha\beta}\,\lambda\left(u,\vec\delta^{\pm}_{(G)}\right)  I_G \,. \label{Tinv}
\end{equation}

The main advantage of this Yangian symmetry formulation is that it reduces to a collection of elementary local exchange relations. In this way, we elucidate how the regular structure of the fishnet graphs translates into their integrability.
The monodromy operator, which is a non-local operator acting on all dual momenta $\{ x \}$ of an $n$-particle fishnet graph $G$, is factorized in the matrix product of the local operators $L_j$ called the Lax operators. The latter satisfy simple local exchange relations with the vertices and propagators constituting the Feynman integal $I_G$. Then the Yangian invariance \re{Tinv} of the fishnet $G$ is verified by applying a sequence of local exchange relations. 

Let us spell out some  details. We define the monodomy operator \re{Texp} as the matrix product of $n$ Lax operators $L_j$, 
\begin{equation}
T\left(u,\vec\delta^{\pm}\right) = L_n[\delta^+_n,\delta^-_n] \cdots L_2[\delta^+_2,\delta^-_2] L_1[\delta^+_1,\delta^-_1] \,, \label{Tmonod}
\end{equation}
where $[\delta^+,\delta^-]\equiv(u+\delta^+,u+\delta^-)$. The Lax operators
$L_j$ are $4\times 4$ matrices whose entries are the conformal generators $J^A_j$ \re{confKx} 
acting on the dual momenta $x_j$, see eq. \re{Texp},
\begin{equation}
L_{\alpha\beta}[\delta^+,\delta^-]= [\delta^+ + \delta^-] \delta_{\alpha\beta} + \mathbb{J}_{\alpha\beta} \,, 
\end{equation}
where $[\delta]\equiv u+\delta$.
The inhomogeneity parameters shift the spectral parameter and specify the conformal weights, $\delta^+ - \delta^- +2 = \Delta$.

\begin{figure}
\begin{equation*}
\hspace*{-26mm}
\begin{array}{c}\scalebox{0.65}{\includegraphics{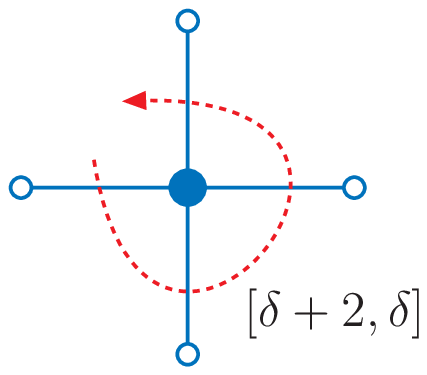}} \end{array} = [\delta+2] 
\begin{array}{c}\scalebox{0.65}{\includegraphics{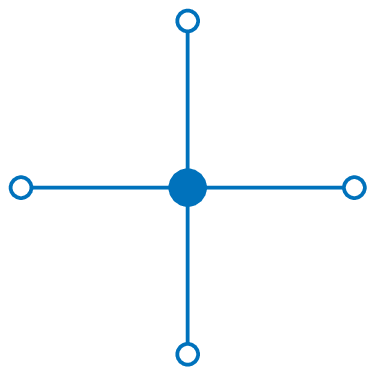}} \end{array}
\hspace{1.5cm}
\begin{array}{c}\scalebox{0.8}{\includegraphics{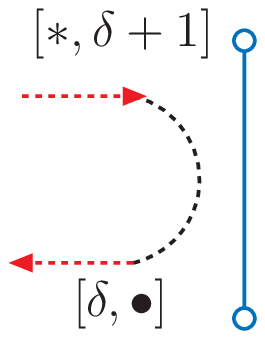}} \end{array} =  
\begin{array}{c}\scalebox{0.8}{\includegraphics{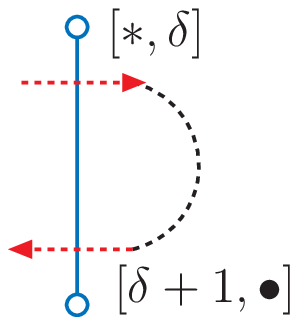}} \end{array}
\end{equation*}
\caption{Local intertwining relations \re{Lf=f} and \re{x2LL=LLx2} satisfied by the Lax operators, propagators and interaction vertices of the dual fishnet graphs.}\label{FigLocIntwRel}
\end{figure}

It is very instructive to represent the Yangian invariance \re{Tinv} graphically as in example in Figure~\ref{FigPi}.
Then the Yangian invariance of fishnets is verified by a sequence of graphical transformations. 
The monodromy can be depicted as a lasso (with $L$'s ordered along the contour) encircling a fishnet graph $G$. Integrating by parts the Lax operator one gets
\begin{equation}
\int d^4 x_0 \, L_0[\delta+2,\delta] f(x_0,\ldots) = [\delta+2]\int d^4 x_0 \, f(x_0,\ldots) \,. \label{Lf=f}
\end{equation} 
Using this relation,
we can deform the lasso contour inside the fishnet graph and entwine the interaction vertex $x_0$ with it, see Figure~\ref{FigLocIntwRel} on the left. Then we pull the lasso inside the bulk of the fishnet due to the intertwining relation for the product of Lax operators and the propagator
\begin{equation}\label{seq}
x_{ij}^{-2} \, L_j[\delta,\bullet] L_i[\ast,\delta+1] = L_j[\delta+1,\bullet]L_i[\ast,\delta]\, x_{ij}^{-2} \,. \label{x2LL=LLx2}
\end{equation}
The latter is depicted in Figure~\ref{FigLocIntwRel}, on the right. 
Eventually, the fishnet graph escapes from the monoromy lasso which turns into the eigenvalue $\lambda$ from \re{Tinv} due to $L[\delta,\delta+2] \cdot 1 = [\delta+2] \cdot \mathds{1}$ that proves \re{Tinv}. For example, we can shrink the lasso in Figure~\ref{FigPi} to a point applying the graphical transformations from Figure~\ref{FigLocIntwRel}.

As we can see, the Yangian invariance takes place only if the inhomogeneities $\vec \delta^{\pm}$ can be adjusted in such a way that each of the local transformations \re{seq} in the sequence holds. This is an exceptional situation which takes place for the fishnets having the regular square lattice structure in the bulk.

\subsection{Yangian symmetry in other fishnet theories}
 
The observation on the Yangian symmetry of the amplitudes in the four-dimensional conformal fishnet model \re{L} allows for numerous generalizations. Adjusting appropriately the double-scaling limit of the $\gamma$-deformed ${\cal N} = 4$ SYM results in an integrable chiral theory of massless scalars and Majorana fermions interacting via Yukawa vertices \cite{Caetano:2016ydc} mentioned after eq.~\re{L}. The corresponding `brick wall' Feynman graphs, which involve fermionic lines, are Yangian invariant \cite{Chicherin:2017frs} with an appropriate spinorial extension of the conformal generators \re{confKx} employed in \re{JJhat}.  

Another direction in the searches for Yangian invariant scalar fishnet graphs consists in varying the space-time dimension $d$,  propagators $(x_{ij}^{-2})^\alpha$, and valency of the interaction vertices. The planar Feynman graphs which are cut out with scissors out of the regular tilings of the plane feature the Yangian symmetry as well \cite{Chicherin:2017frs}. Apart from the square lattice of the $4d$ fishnet theory \re{L}, there are triangular and hexagonal regular tilings corresponding to $\varphi^6$ interactions in $d=3$ with $|x_{ij}|^{-1}$ propagators and $\varphi^3$ in $d=6$ with $|x_{ij}|^{-4}$ propagators, respectively. The underlying integrable $3d$ fishnet theory is a double-scaling of the $\gamma$-deformed ABJM theory \cite{Caetano:2016ydc} and the $6d$ fishnet theory is studied in \cite{Mamroud:2017uyz}. 

Leaving out the underlying Lagrangian theory, we can start looking for Yangian invariant fishnet graphs formed by scalar propagators $(x_{ij}^{-2})^{\alpha_{ij}}$ with noninteger powers $\alpha_{ij} = \alpha_{ji}$ and $d$-dimensional interaction vertices $\int d^d x_i$. In order to respect the conformal part $J^A$ of the symmetry, see eq. \re{confKx}, the  powers of the propagators joining a common vertex $x_i$ should be balanced $\sum_{j} \alpha_{ij} = d$. If the propagator powers are balanced at each vertex of a planar graph $G$ cut out with scissors from a regular tiling of the plane (triangular, square or hexagonal) then $I_G$ is Yangian invariant.

The fishnet graphs considered so far involved solely massless scalar propagators. Fishnets with massive propagators are compatible with the Yangian symmetry as well. The relevant set of graphs is suggested by the massive four-dimensional bi-scalar Lagrangian \cite{Loebbert:2020tje,Karananas:2019fox}. In the leading color limit, the massive propagators are pushed onto the boundary of the fishnet graph (and they have the form $(x_{ij}^2 + m_j^2)^{-1}$ where $x_i$ is an internal dual momenta and $x_j$ is an external dual momenta) and all interior propagators of the fishnet graph are massless. The Yangian invariance conjecturally holds for $d$-dimensional fishnet graphs with arbitrary but balanced at each vertex propagator powers cut out of the regular tilings of the plane with massless propagators in the bulk and massive propagators on the boundary \cite{Loebbert:2020hxk}. The masses are introduced in the Yangian generators $\widehat{J}^A$ \re{JJhat} by extending the conformal generators $J^{A} \to J^{(m)\,A}$ from \re{confKx}. One starts with the massless conformal generators in $d+1$ dimensions \re{confKx} and identifies the mass with the $(d+1)$-th direction, $x^{d+1} = m$. This does not change $P_\mu$ and $L_{\mu\nu}$ but alters the boost and dilatation \cite{Alday:2009zm}, 
\begin{eqnarray}\notag
&& D^{(m)} = D - i m \pa_m \,,\qquad 
\\
&& K^{(m)}_\mu = K_\mu -2 i x_\mu m \pa_m + i m^2 \pa_\mu \,.
\end{eqnarray}
The formulation of the massless Yangian symmetry as the closure of two independent conformal symmetries carries over to the massive setting \cite{Loebbert:2020hxk}. The relevant massive extension $\bar{J}^A \to  \bar{J}^{(m)\,A}$ of the second conformal symmetry, namely of the momentum space conformal generators \re{confKp}, is suggested by the equivalence
$\widehat{P}^{(m)}_\mu \sim \bar{K}^{(m)}_\mu$.

\subsection{Yangian symmetry on-shell} \label{sec_Yang-on-shell}

So far we have discussed the Yangian symmetry of the off-shell 'amplitudes' $I_G$ assuming that all external momenta are off-shell, $p_i^2 \neq 0$ or equivalently $x_{i\,i+1}^2 \neq 0$. Now we address the Yangian symmetry of their on-shell counterparts ${\cal I}_G$ which are true single-trace leading-color amplitudes in the conformal  fishnet theory. The dual conformal symmetry, which is formed by the level-zero generators $J^A$ \re{confKx}, is exact. However, the level-one generators $\widehat{J}^A$ \re{JJhat} could acquire the anomaly, which originates from noncommutativity of the symmetry generators and the on-shell limit,
\begin{equation}
\lim_{p_i^2 \to 0} \widehat{J}^A\, I_G \neq \widehat{J}^A\, \lim_{p_i^2 \to 0} I_G \,. \label{non-comm}
\end{equation}
Despite of the fact that the fishnet graphs with  inflowing light-like momenta $p_i^2 = 0$ are finite, their Yangian symmetry is anomalous. The anomaly originates from those vertices of fishnet graphs where a single on-shell momentum flows in (e.g. momenta $p_1$ and $p_6$ in the two-loop graph in Figure~\ref{FigCross} on the right). We show below that the anomaly comes from integration over loop momenta that are collinear to this external light-like momentum.

We outline the anomaly mechanism for the conformal fishnets \re{L} using the dual momenta variables and relying on the RTT-formulation of the Yangian algebra. Let us consider the asymptotic expansion of the Yangian invariant off-shell fishnet $I_G$ with the inflowing momenta approaching the light-cone ($p_i^2 \equiv x_{ii+1}^2 \to 0$),
\smallskip

\begin{equation}
I_G = {\cal I}_G + \sum_i p_{i}^2 \log(p_{i}^2)\,  {\cal I}_G^{[i]} + \sum_{i < j} p_{i}^2 \log(p_{i}^2) p_{j}^2 \log(p_{i}^2)\,  {\cal I}_G^{[ij]} +\ldots \label{IGasymp}
\end{equation}
where the on-shell fishnet ${\cal I}_G$ as well as the coefficients ${\cal I}_G^{[i]}, {\cal I}_G^{[ij]},\ldots$ depend only on the on-shell kinematics. We keep only $p_i^2 \log(p_i^2)$ terms in the expansion \re{IGasymp}, since only they contribute to the anomaly. Indeed, the factor $L_{i+1} L_i$ of the monodromy matrix \re{Tmonod} acting on  $p_{i}^2\log(p_{i}^2)$ gives a nonvanishing contribution at $p_i^2 \to 0$,
\begin{eqnarray}
& \left. L_{i+1}[\delta^+_{i+1},\delta^-_{i+1}] L_{i}[\delta^+_{i},\delta^-_{i}]\cdot p_{i}^2\log(p_{i}^2) \right|_{p_i^2 \to 0} \notag\\ 
& = 
\left( \begin{array}{cc}  \bm{x}_{i i+1} \tilde{\bm{x}}_i  & -\bm{x}_{ii+1} \\
\tilde{\bm{x}}_{i+1} \bm{x}_{ii+1} \tilde{\bm{x}}_{i}  & - \tilde{\bm{x}}_{i+1} \bm{x}_{i i+1}
\end{array} \right) \equiv M_{i+1\,i} \;, \quad \hbox{at} \quad \delta^-_i - \delta^+_{i+1} = 1 \,, \label{Mmat}
\end{eqnarray}
where $\bm{x}:= \textup{i} \sigma^\mu x_\mu$ and $\tilde{\bm{x}} := -\textup{i} \bar\sigma^\mu x_\mu$ are defined with the help of $2\times 2$ Pauli matrices $\sigma^\mu=(1,\vec \sigma)$ and $\bar\sigma^\mu=(1,-\vec \sigma)$. The  above equation clarifies the noncommutativity in \re{non-comm}.

The exact Yangian invariance \re{Tinv} of the off-shell `amplitude' $I_G$ boils down to the anomalous Yangian invariance of the amplitude ${\cal I}_G$,  
\begin{eqnarray}
 T_{\alpha\beta}\left(u,\vec\delta^{\pm}_{(G)}\right) {\cal I}_G && - \delta_{\alpha\beta}\,\lambda\left(u,\vec\delta^{\pm}_{(G)}\right)  {\cal I}_G \notag\\ 
 &&= - \sum_{k \geq 1}\sum_{i_1,\ldots,i_k} T^{[i_1,\ldots,i_k]}_{\alpha\beta}\left(u,\vec\delta^{\pm}_{(G)}\right) {\cal I}^{[i_1,\ldots,i_k]}_G \,. \label{TinvAnom}
\end{eqnarray}
To obtain this relation, one acts with the monodromy matrix \re{Tmonod} on the expansion \re{IGasymp} and takes into account \re{Mmat}.
The right-hand side of the anomaly equation \re{TinvAnom} involves the anomalous monodromy matrix $T^{[i_1,i_2,\ldots,i_k]}$. Here the labels are such that $|i_l - i_{l+1}|>1$, $l=1,\ldots,k-1$. It is obtained from the monodromy matrix \re{Tmonod} by replacing $k$ pairs of neighboring Lax operators with $M$-matrices \re{Mmat}, $L_{i_l+1}L_{i_l} \to M_{i_l+1\, i_l}$, $l=1,\ldots,k$. Let us note that the relation on the inhomogeneity parameters $\delta^-_{i_l} - \delta^+_{i_l+1} = 1$ from \re{Mmat} is satisfied (see e.g. Figure~\ref{FigPi}). 
The anomalous monodromy matrix $T^{[i_1,\ldots,i_k]}$  is a polynomial in the spectral parameter $u$ of degree $n-2k$ (with $k\ge 1$). As a consequence, it does not contribute to $O(u^{n-1})$ terms on both sides of \re{TinvAnom} and does not affect the level-zero symmetry generated by the operators $\mathbb J$ defined in \re{Texp}, as expected.

The coefficients $ {\cal I}^{[i_1,\ldots,i_k]}_G$ of the asymptotic expansion \re{IGasymp} are simpler to evaluate than the amplitude ${\cal I}_G$. They are given by the fishnet graphs $G$ in which $2k$ loop momenta are collinear with the adjacent external on-shell momenta (see eq. \re{Cdef} and \re{prodC}). This reduces the order of loop integrations by $2k$. In order to see how it works in practice, we single out three propagators from an on-shell fishnet $G$ which are adjacent to the external dual momenta $x_i$ and $x_{i+1}$,
\begin{equation}
{\cal I}_G =  \int \frac{d^4 x_0 d^4 x_{0'}}{(2\pi^2)^2} \frac{f_G(x_0,x_{0'},\ldots)}{x_{i 0}^2 x_{00'}^2 x_{i+1 0'}^2} = 
\begin{array}{c}\scalebox{0.6}{\includegraphics{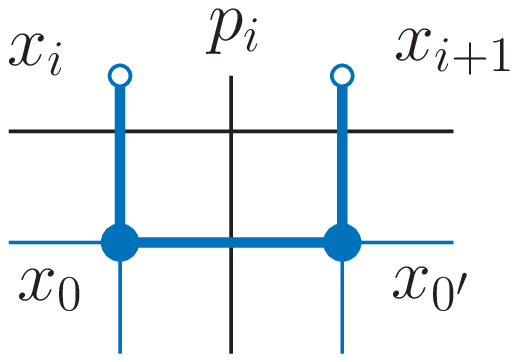}} \end{array}
\,,
\end{equation}
where $f_G$ is the remaining part of the fishnet graph. We define the operation ${\bf C}_{i\,i+1}$ which localizes the three propagators on the collinear configuration $x_{i0} =\alpha p_i$, $x_{0' i+1} = \beta p_i$ and $x_{00'} = (1-\alpha-\beta)p_i$ as follows \cite{Chicherin:2017bxc},
\begin{eqnarray}\notag
{\bf C}_{i\,i+1}\,  {\cal I}_G && := \int\limits_{\alpha,\beta \geq 0 \atop \alpha + \beta \leq 1 } d\alpha d\beta \, f_G(x_i - \alpha x_{i i+1},x_{i+1} + \beta x_{i i+1},\ldots) 
\\
&& = \begin{array}{c}\scalebox{0.6}{\includegraphics{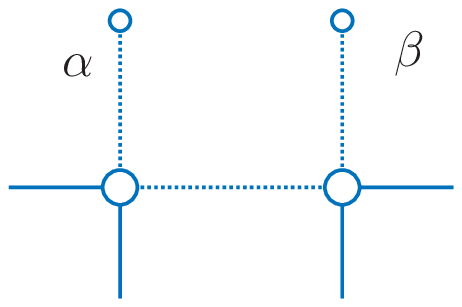}} \end{array} \label{Cdef}
\end{eqnarray}
The coefficients of the asymptotics expansion \re{IGasymp} are obtained by the multiple action of ${\bf C}$'s operations (the ordering is irrelevant)
\begin{equation}
 {\cal I}^{[i_1,\ldots,i_k]}_G = \prod_{l=1}^k {\bf C}_{i_l\,i_l+1}\cdot  {\cal I}_G \,. \label{prodC}
\end{equation}
Thus, if ${\cal I}_G$ is an $\ell$-loop Feynman graph then ${\cal I}^{[i_1,\ldots,i_k]}_G$ is a $(\ell-2k)$-loop Feynman graph integrated over $2k$ auxiliary parameters.

The Yangian anomaly is absent for the one-loop box ${\cal I}_{\displaystyle +}$ \re{Icross}. Indeed, after putting the external momenta on-shell the corners of the box in Figure~\ref{FigCross} are still off-shell. The two-loop train track graph ${\cal I}_{+\!\!+}$ ($\ell = 2$, see Figure~\ref{FigPi}) with ten external on-shell momenta $p_i^2 = 0$ provides the simplest example of a nontrivial Yangian anomaly. In this case the anomaly is attributed to the loop momenta configurations which are collinear either with $p_1^\mu$ or $p_6^\mu$; and the anomaly equals to $-T^{[1]}{\cal I}^{[1]}_{+\!\!+}-T^{[6]}{\cal I}^{[6]}_{+\!\!+}$ where both ${\cal I}^{[1]}_{+\!\!+} = {\mathbf C}_{12} {\cal I}_{+\!\!+}$ and ${\cal I}^{[6]}_{+\!\!+} {\mathbf C}_{67}{\cal I}_{+\!\!+}$ are rational functions integrated over two auxiliary parameters, see \re{Cdef}.

The Yangian symmetry in the momentum representation \re{JhatMoment} is anomalous as well for the on-shell conformal fishnets in $d=4$. Indeed, the momentum space conformal boost $\bar{K}^\mu$ \re{confKp} is anomalous \cite{Chicherin:2017bxc}, so we conclude that even the level-zero part of the momentum Yangian is anomalous. The same observation concerns the momentum representation fishnet graphs with scalar $(\phi^3)_{6d}$ and $(\phi^6)_{3d}$ interaction vertices, as well as four-dimensional fishnets with Yukawa interactions. In all these cases, the Yangian anomaly is localized on the collinear configurations of the loop momenta \cite{Chicherin:2017bxc}.

\section{Conclusions}

Scattering amplitudes contain a wealth of information about the dynamics of the underlying quantum field theory. Finding them exactly would be equivalent to solving the theory. Existing techniques allow us to compute scattering amplitudes in interacting four-dimensional field theories only to the few first orders in the weak coupling expansion. A considerable amount of progress  has been recently achieved in the maximally supersymmetric ${\cal N}=4$ SYM theory. A powerful arsenal of integrability based methods has been developed to facilitate calculations of the scattering amplitudes in this theory both at weak and at strong coupling. In spite of this, the weak coupling expansion in ${\cal N}=4$ SYM involves a multitude of Feynman diagrams. This obscures the simplicity of the amplitudes and renders the conventional Feynman diagram approach rather impractical. 

The conformal fishnet theory offers a remarkable example of a nontrivial interacting nonsupersymmetric field theory.  
The scattering amplitudes in this theory have the same integrability properties as in planar ${\cal N}=4$ SYM and, at the same time, have much simpler Feynman diagram expansion. This allows one to obtain a compact representation for the simplest four-particle amplitude and to explain integrability of multi-particle scattering amplitudes using the properties of individual Feynman diagrams. In a broader context, the conformal fishnet theory helps us to better understand the origin of integrability of planar ${\cal N}=4$ SYM and to develop further the integrability approach to computing the scattering amplitudes.

 In this review, we restricted our consideration to the scattering amplitudes in the conformal fishnet theory.  Another very active area of research 
concerns applications of integrability methods to off-shell quantities in this theory, e.g. correlation functions of local operators \cite{Gromov:2018hut,Kazakov:2018gcy}.  In the weak-coupling expansion of a generic correlation function, the contributing planar Feynman diagrams are closely related to those shown in Figure~\ref{fig:cross}. Namely, they can be obtained from the amplitude graphs by taking the external legs to be off-shell and by attaching them to external space-time points.  

The regular square lattices of an arbitrary size is a special class of such Feynman integrals. Attaching the external legs on each side of the square to four distinct points we obtain a family of  off-shell four-point  multi-loop Feynman integrals \cite{Basso:2017jwq}, which generalizes the well-known   ladder integrals \cite{Usyukina:1993ch}.  This family is a remarkable example of Feynman integrals with nontrivial kinematic dependence known in a closed form at any loop order. Applying integrability techniques supplemented with the generalized Steinmann relations, which take into account nontrivial analytical properties of the Feynman integrals,
 one can derive a closed form determinant representation of these integrals in terms of ladder integrals \cite{Basso:2017jwq,Derkachov:2019tzo,Derkachov:2020zvv,Derkachov:2021rrf}. For further development in this field, we refer the interested readers to Refs.~\cite{Basso:2019xay,Basso:2021omx}.
 
\section*{Acknowledgments}

We are grateful to Tristan Mc Loughlin for carefully reading the manuscript and for useful remarks.
This work  was supported  by the European Union's Horizon 2020 research and innovation programme under the Marie Sk\l{}odowska-Curie grant agreement No.~764850 {\it ``\href{https://sagex.org}{SAGEX}''}. We also acknowledge support from 
the French National Agency for Research in the framework of the \textit{Investissements d'avenir} program (ANR-15-IDEX-02)
and grant ANR-17-CE31-0001-01. 
\\
 

\providecommand{\newblock}{}

\end{document}